\documentclass[]{spie}  

\usepackage{amsmath,amsfonts,amssymb}
\usepackage{graphicx}
\usepackage[colorlinks=true, allcolors=blue]{hyperref}
\usepackage{colortbl}
\usepackage{subcaption}
\usepackage[rightcaption]{sidecap}
\usepackage{xspace}    

\newcommand{\micron}{\textmu{}m\xspace}   

\title{SPICA-FT: The new fringe tracker of the CHARA array}

\author[a]{Cyril Pannetier, Philippe Berio, Denis Mourard, Sylvain Rousseau, Fatme Allouche, Julien Dejonghe, Christophe Bailet, Daniel Lecron}
\author[b]{Fr\'ed\'eric Cassaing}
\author[c]{Jean-Baptiste Le Bouquin, Karine Perraut}
\author[d]{John D. Monnier}
\author[e]{Narsireddy Anugu, Theo ten Brummelaar}

\affil[a]{Laboratoire Lagrange, Universit\'e C\^ote d’Azur, Observatoire de la C\^ote d’Azur, CNRS, Boulevard de l’Observatoire, CS 34229, 06304
Nice Cedex 4, France}
\affil[b]{ONERA, DOTA, Paris Saclay University, France}
\affil[c]{Univ. Grenoble Alpes, CNRS, IPAG, 38000 Grenoble, France}
\affil[d]{University of Michigan, Ann Arbor, MI 48109, USA}
\affil[e]{CHARA Array, Georgia State University, Atlanta, GA 30302, USA}

\begin{document} 

\graphicspath{ {images/} }

\maketitle

\begin{abstract}
SPICA-FT is part of the CHARA/SPICA instrument which combines a visible 6T fibered instrument (SPICA-VIS) with a H-band 6T fringe sensor. SPICA-FT is a pairwise ABCD integrated optics combiner. The chip is installed in the MIRC-X instrument. The MIRC-X spectrograph could be fed either by the classical 6T fibered combiner or by the SPICA-FT integrated optics combiner. SPICA-FT also integrates a dedicated fringe tracking software, called the opd-controller communicating with the main delay line through a dedicated channel. We present the design of the integrated optics chip, its implementation in MIRC-X and the software architecture of the group-delay and phase-delay control loops. The final integrated optics chip and the software have been fully characterized in the laboratory. First on-sky tests of the integrated optics combiner began in 2020. We continue the on-sky tests of the whole system (combiner + software) in Spring and Summer 2022. We present the main results, and we deduce the preliminary performance of SPICA-FT.  
\end{abstract}

\keywords{Interferometry, Fringe tracker}

\section{Introduction}
Unlike VLTI with GRAVITY-FT\cite{lacour2019} , the fringe tracker of the 4-beam spectro-interferometer GRAVITY\cite{gravity_collaboration_first_2017} , the CHARA array was not equipped with such an instrument. On VLTI, the need for a fringe-tracker was in particular justified by the scientific objective of GRAVITY to precisely measure the dynamical processes around the central black hole of the Milky Way, Sgr A*\cite{gravity_collaboration_detection_2018} . In the context of the SPICA instrument\cite{mourard2017} , we developed the fringe tracker SPICA-FT. It operates simultaneously with SPICA in order to push its sensitivity. Our goal is to stabilize the fringes over 100–200 ms with an accuracy of the order of 100 nm. Even if SPICA-FT was developed for SPICA, it could be used to stabilize fringes for other beam combiners of CHARA. It could be considered as the CHARA fringe tracker.

SPICA-FT is inspired from GRAVITY-FT. It recombines up to 6 telescopes in H band with a pairwise ABCD encoding in an integrated optics (IO) chip. The IO chip is installed in the MIRC-X instrument\cite{anugu_mirc-x_2020} already in operation at CHARA and should provide a higher sensitivity\cite{minardi_beam_2016} than the all-in-one optical setup. Such integrated optics have first been proposed for astronomical purpose in 1999\cite{malbet1999} . The chip output is placed at the entrance of the MIRC-X spectrograph. SPICA-FT could be operated at several spectral resolution, the lowest spectral resolution (R=22) giving the best sensitivity.

In addition to the optical module, we developed a dedicated fringe tracker software. It is fully integrated in the MIRC-X control software. It is also inspired from the GRAVITY fringe tracker software. It allows Group Delay or Phase Delay tracking. The algorithm follows the formalism presented in Ref.~\citenum{lacour2019}. Its architecture has been defined in such a way that it could be used with several beam combiners: SPICA-FT of course but also with the standard MIRC-X combiner or with the MYSTIC combiner\cite{monnier2018} .

A first version of the SPICA-FT chip has been installed and tested on sky in 2020. We obtained fringes with 5 telescopes (one telescope was not available at the time). Then we worked on the optimization of the main functions of the IO Chip and on the development of the fringe tracker software. In May 2022, we tested the final version of the software on sky with the standard MIRC-X combiner. The final IO chip will be installed during Summer of 2022.

In this paper, we present the IO combiner and its performance in section \ref{IOchip}, the fringe tracker software in section \ref{ftsoft} and the results obtained on sky in May 2022 in section \ref{results}.

\section{Beam Combiner}
\label{IOchip}
\subsection{Design and Manufacturing of the IO chip }

\begin{figure}[ht]
\begin{center}
\includegraphics[width=0.9\textwidth]{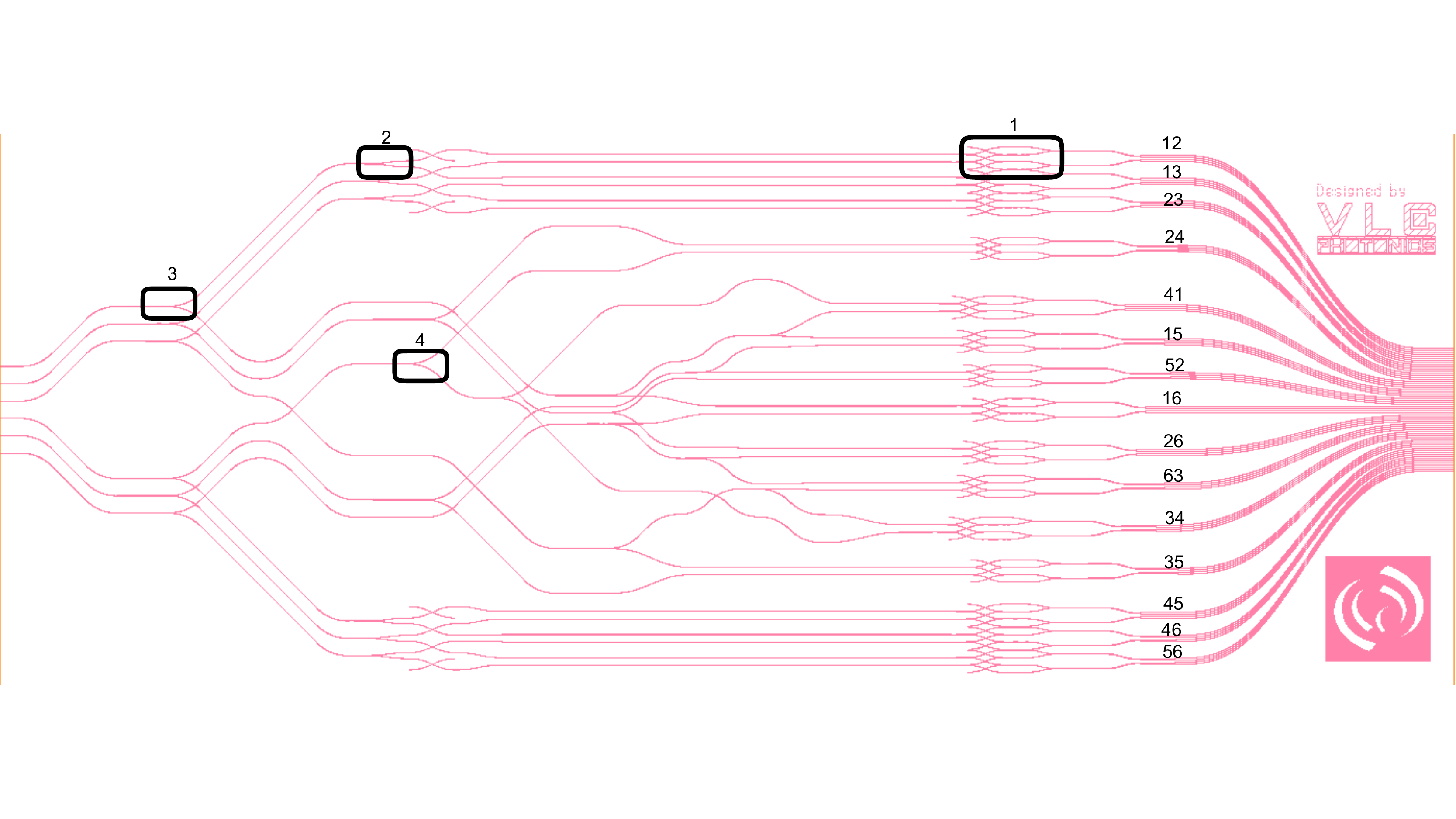}
\caption{Design (VLC-OCA) of the SPICA-FT IO beam combiner. Each beam is divided in 5 equal parts thanks to a smart succession of 60/40 (3), 50/50 (2), and 66/33 (4) couplers. The 15 possible pairs are combined inside a ABCD circuit (1), sampling the fringes with fixed phase shifts of $0$, $\pi/2$, $\pi$, and $3\pi/2$. This permits the instantaneous measurements of the phase-delay. Optical paths are equalized to avoid any bias in closure phase measurements.}
\label{labelchip}
\end{center}
\end{figure}

In October 2017, we consulted at least five different companies for the fabrication of an integrated optics chip able to combine, following the ABCD scheme, 6 beams over the H band. It is a pairwise scheme (Figure \ref{labelchip}) leading to 15 baselines and $15\times 4=60$ outputs. The initial scheme was based on the design proposed by Pierre Labeye in his PhD manuscript~\cite{labey2008}. We have defined the following specifications:
\begin{itemize}
\item operating wavelength range: 1.5 to 1.8~\micron ; 6 inputs, 15 baselines with ABCD encoding,
\item single mode waveguides over the H band with a Numerical Aperture (at 1/e²) between 0.13 and 0.15,
\item Phase shifts 90°+-10° over the spectral band,
\item Throughput larger than 60\% over the H band,
\item Contrast level larger than 95\% in polarized light,
\item Flux balancing between two entrance waveguides with a tolerance of 15\%, Flux cross-talk below 0.5\%.
\end{itemize}

In December 2017, we started a feasibility study with the VLC Photonics company on the technology of Planar Lighwave Circuit, with 1.8\% of difference of refractive index between the core guiding structure and the cladding layers (technology of doped silica). The detailed study was then started in December 2018 and we received our first chips in May 2019. Unfortunately many of these chips were corrupted by large defaults during the manufacturing process (large bubbles and poor polishing of the entrance and exit faces). After analysis, we identified additional unbalances in the 66/33 and in the 50/50 splitter functions. However in terms of transmission (55\%) and in terms of phase shift quality the best chip was very close to the specifications.

A second iteration for the manufacturing process was done after a series of meetings with the company and the foundry to improve the cleaning process before the fabrication of our large chips (82x35$mm$) whereas the foundry was mainly manufacturing a large number of small chips where defaults just lead to rejection of a few percentage of the fabrication. In our case of fabrication of a small number of chips (considering the cost and the size of a wafer) the probability of pollution should be reduced as much as possible. With this in mind, the foundry fabricated a second series of chips where the splitter functions have been redesigned. This second series was of much better quality but we ended with still a poor flux balancing because of a residual default in the 66/33 splitter function. Moreover our characterization demonstrated that the design made by the company was not exempt of internal closure phase. This second generation was tested on sky in January 2020. After new exchanges with the VLC Photonics company, we obtained, with the third generation received in October 2021, excellent results with low internal closure phases, good balancing, and overall a good match with our specifications.

The beams coming from the telescope are injected into the IO beam combiner thanks to optical fibers. We selected polarization maintaining single mode fibers (HB1000 6/125). The fibers were assembled by Leukos company into a silicon V-groove array. Slow polarization axes are guaranteed by Leukos to be perpendicular to the main axis of the front face of the V-groove to within $\pm 3\deg$. The fibers were then length-matched ($\pm$ 0.5~mm) and connectorized (FC/PC). The V-groove is then glued to the entrance of the IO beam combiner.  During the gluing process, we fed the fibers with a source and we measured the flux at the IO beam combiner outputs to check the matching between the fibre core and IO waveguides. When the matching is maximized, the glue is UV polymerized to “freeze” the adjustment.

\begin{figure}
\begin{center}
\includegraphics[width=0.5\textwidth]{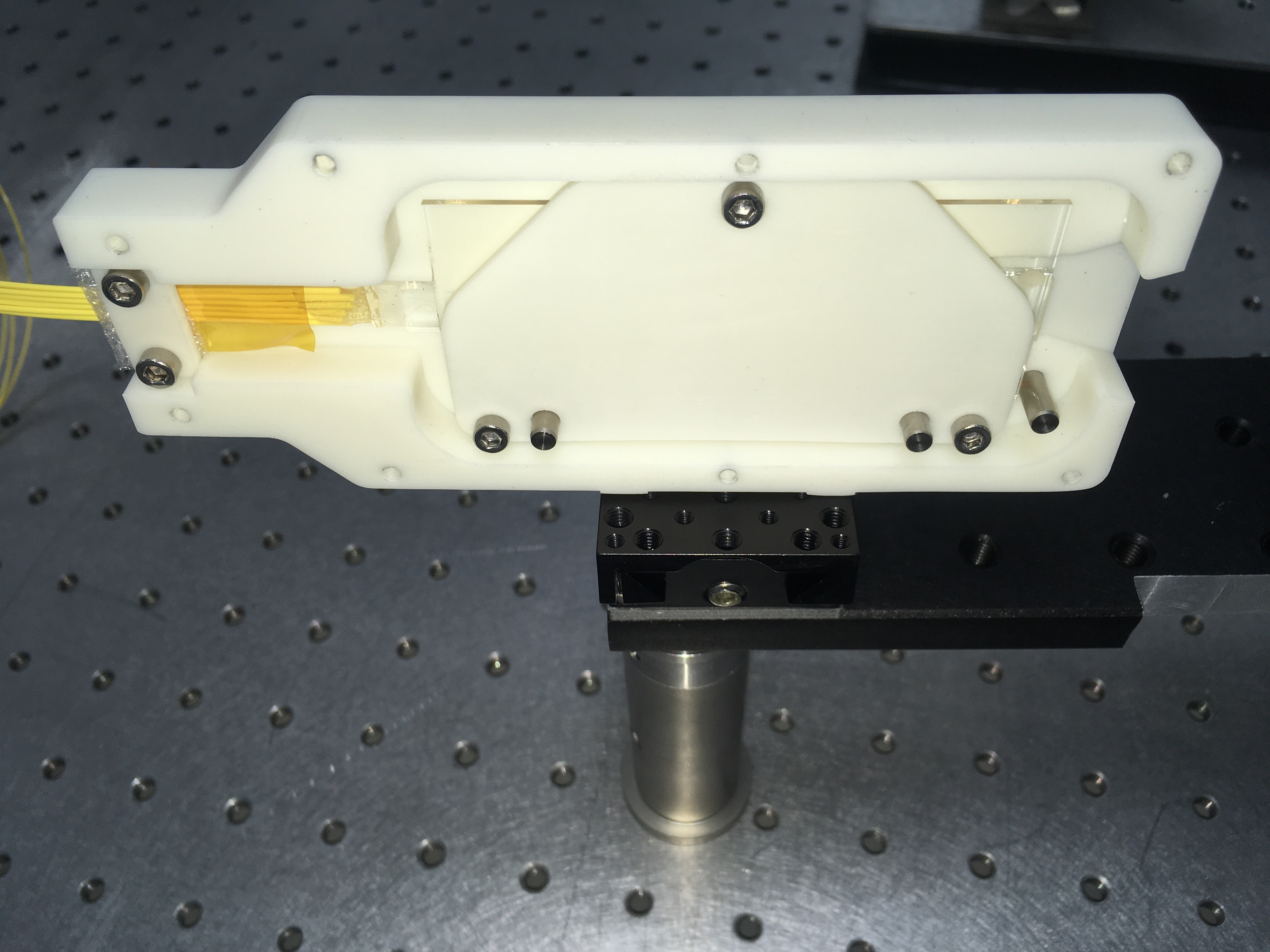}
\caption{Picture of the whole V-groove/chip/MLA. The V-groove is on the left of the IO beam combiner and the MLA is on the right.}
\label{figchipmla} 
\end{center}
\end{figure}

At the output of the IO beam combiner, we glued a micro lens array (MLA). It has been specified in order to feed the MIRCx spectrograph with beams with the correct f-number. This fused silica MLA was manufactured by Advanced Microoptic Systems GmbH company. It is a 1D-array of refractive positive microlenses with a pitch of 120~\micron, a lens diameter of 120~\micron, a lens clear aperture of 100~\micron and a curvature radius of 130~\micron. The MLA thickness is 0.3~mm and it is anti-reflection coated on both sides.\\
Figure \ref{figchipmla} shows a picture of the whole V-groove/chip/MLA inside its mount.  

\subsection{Performance}
We used laboratory testbenchs to check the main characteristics of all manufactured IO chips. The
global transmission has been measured on a dedicated testbench in Grenoble while we developed a 6T fringe tracker testbench in Nice in order to characterize the interferometric properties of the IO chips.

First, we measured the global throughput of all the manufactured assemblies. We found an average throughput at 1.62~\micron higher than 55\% which is close to our specifications.

To characterize the interferometric properties of the IO chip, we used the P2VM formalism\cite{tatulli2007}. The P2VM is measured following the procedure described in Lapeyrere et al 2014\cite{lapeyrere2014}. The P2VM characterizes the relations of the photometry, the coherence, and the phase between the 6 inputs and
the 60 outputs of the IO beam-combiner. The $\kappa\_$matrix contains the contribution of each beam to each fringe pattern. It could be extracted from the P2VM. In a perfect case, a fifth of the flux of each beam contributes to each fringe pattern. With our best chip, we found a contribution of $20\pm3\%$ instead of $20\%$ and we measured an averaged cross-talk of 
$-0.1\pm0.1\%$. The corresponding $\kappa\_$matrix is presented in Table \ref{table-kappa}.

\begin{table}[ht]
\caption{$\kappa\_$matrix: contribution (in \%) of each beam to each fringe pattern} 
\label{table-kappa}
\begin{center}       
\begin{tabular}{|c|c|c|c|c|c|c|} 
\hline
 & Beam 1 & Beam 2 & Beam 3 & Beam 4 & Beam 5 & Beam 6  \\
\hline
\hline
Base 12 & \cellcolor[gray]{0.8}19.3 &	\cellcolor[gray]{0.8}22.7 &	0.0 &	0.2 &	0.1 &	0.1 \\
\hline
Base 13 & \cellcolor[gray]{0.8}22.7 &	-0.2 &	\cellcolor[gray]{0.8}17.4 &	0.1 &	0.1 &	0.1 \\
\hline
Base 23 & -0.1 &	\cellcolor[gray]{0.8}22.7 &	\cellcolor[gray]{0.8}23.1 &	-0.3 &	0.0 &	0.0 \\
\hline
Base 24 & 0.0 &	\cellcolor[gray]{0.8}20.3 &	-0.3 &	\cellcolor[gray]{0.8}18.8 &	-0.3 &	-0.1 \\
\hline
Base 14 & \cellcolor[gray]{0.8}17.6 &	-0.1 &	-0.2 &	\cellcolor[gray]{0.8}18.0 &	-0.3 &	-0.3 \\
\hline
Base 15 & \cellcolor[gray]{0.8}18.2 &	0.0 &	-0.2 &	-0.1 &	\cellcolor[gray]{0.8}16.3 &	-0.3 \\
\hline
Base 25 & -0.1 &	\cellcolor[gray]{0.8}16.0 &	-0.1 &	0.0 &	\cellcolor[gray]{0.8}16.7 &	-0.1\\
\hline
Base 16 & \cellcolor[gray]{0.8}22.5 &	-0.1 &	-0.2 &	-0.3 &	-0.1 &	\cellcolor[gray]{0.8}21.1\\
\hline
Base 26 & -0.1 &	\cellcolor[gray]{0.8}19.9 &	-0.1 &	-0.4 &	-0.2 &	\cellcolor[gray]{0.8}19.2\\
\hline
Base 36 & 0.0 &	-0.1 &	\cellcolor[gray]{0.8}18.4 &	-0.1 &	-0.3 &	\cellcolor[gray]{0.8}16.5\\
\hline
Base 34 & 0.0 &	0.0 &	\cellcolor[gray]{0.8}19.2 &	\cellcolor[gray]{0.8}16.9 &	-0.1 &	-0.2\\
\hline
Base 35 & 0.0 &	-0.1 &	\cellcolor[gray]{0.8}23.2 &	0.0 &	\cellcolor[gray]{0.8}21.3 &	-0.2\\
\hline
Base 45 & 0.0 &	0.0 &	-0.1 &	\cellcolor[gray]{0.8}25.1 &	\cellcolor[gray]{0.8}23.1 &	-0.1\\
\hline
Base 46 & 0.0 &	0.0 &	0.0 &	\cellcolor[gray]{0.8}22.2 &	-0.1 &	\cellcolor[gray]{0.8}27.6\\
\hline
Base 56 & 0.0 &	0.0 &	0.0 &	0.0 &	\cellcolor[gray]{0.8}23.7 &	\cellcolor[gray]{0.8}16.6\\
\hline 
\end{tabular}
\end{center}
\end{table}

\begin{figure} [ht]
\begin{subfigure}[b]{0.45\linewidth} \centering
\includegraphics[width=\textwidth]{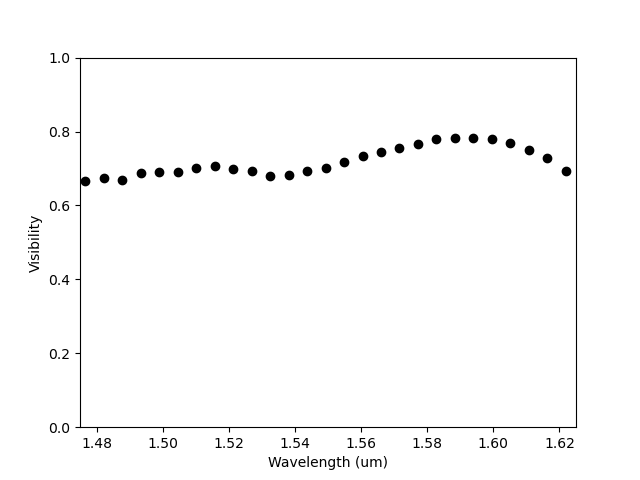}
\end{subfigure} \hfill
\begin{subfigure}[b]{0.45\linewidth} \centering
\includegraphics[width=\textwidth]{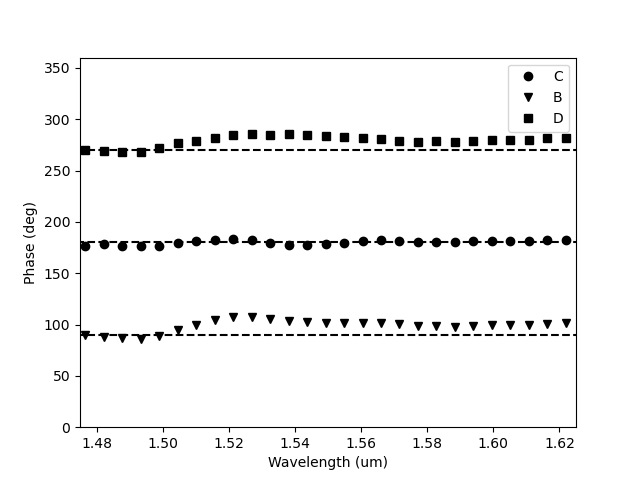}
\end{subfigure}
\caption{Instrumental Contrast and Phase Shifts for the baseline 16.}
\label{figshift} 
\end{figure}

The P2VM provides also information on the instrumental visibility and on the phase shifts between the ABCD outputs. We found an instrumental visibility (averaged over 15 baselines and over 27 spectral channels between 1.45 and 1.65~\micron) of $74\pm13\%$. It should be noted that these visibility measurements were done in natural light and without any compensation of the birefringence of the fibers. This last effect is corrected when the component is installed inside MIRCx. For all baselines and all the wavelengths, the ABCD outputs should be in quadrature (i.e. phase differences around $90\deg$). We found phases (averaged over 15 baselines and over 27 spectral channels between 1.45 and 1.65~\micron) of output B, C and D with respect to output A equal to $84.1\pm16.0\deg$, $181.3\pm2.7\deg$ and $264.4\pm16.1\deg$ respectively. 
As an example, Figure \ref{figshift} shows the instrumental contrast and the phase shifts of the BCD outputs versus the A output for the Baseline 26.

\subsection{Installation in MIRCX}
\begin{figure} [ht]
\centering
\includegraphics[width=0.8\textwidth]{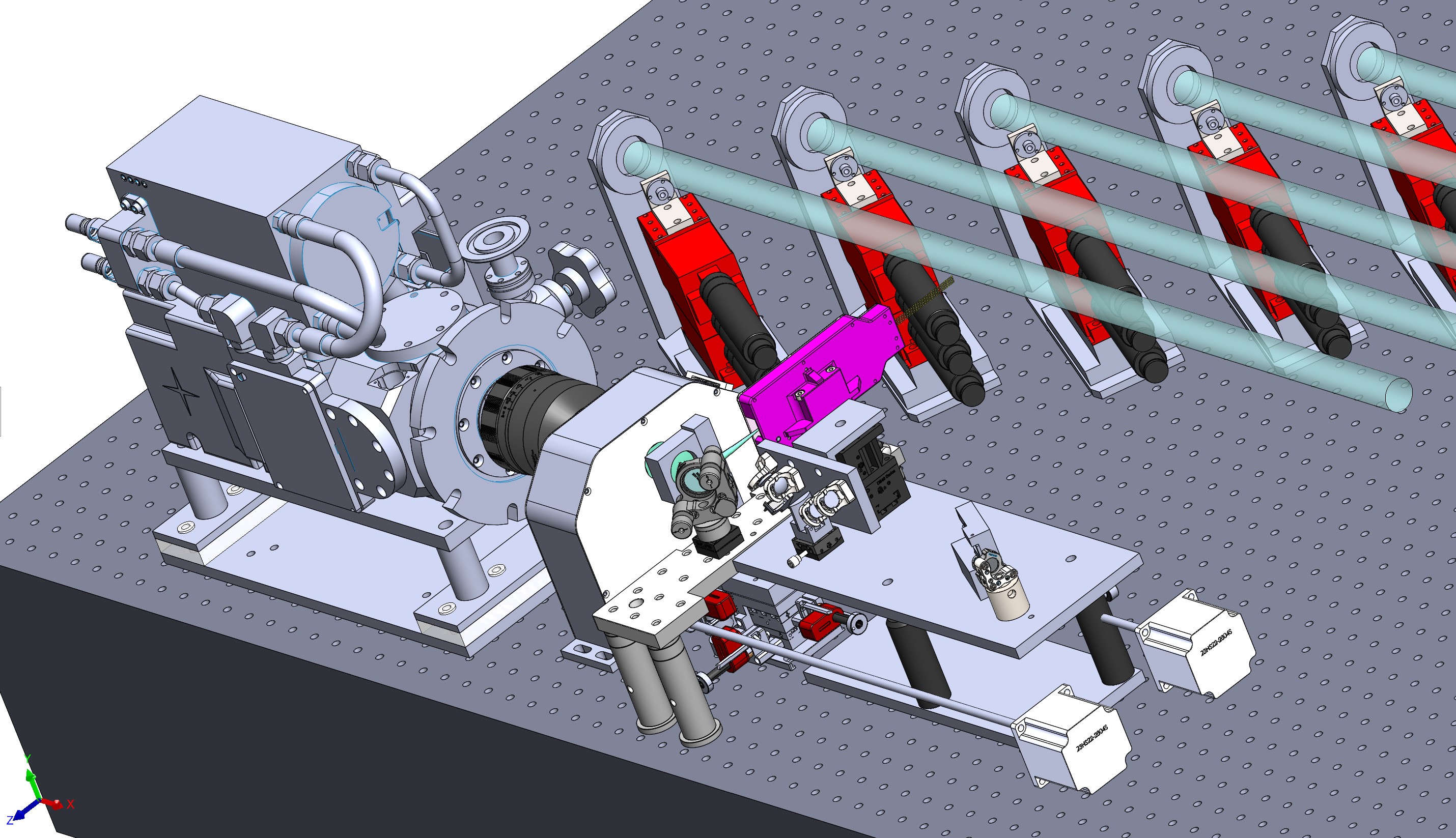}
\caption{3D view of the MIRCx focal instrument with the SPICA-FT IO beam combiner (in pink).}
\label{figmircx} 
\end{figure}

The SPICA-FT IO beam combiner is installed at the entrance of the MIRCx spectrograph. In figure \ref{figmircx}, the MIRCx detector is represented on the left, the standard MIRCx beam combiner is on the bottom right. The spectrograph is located between these two modules (dispersive optics wheel and collimator). A removable folding mirror could be used to feed the spectrograph with the SPICA-FT beam combiner (pink module in figure \ref{figmircx}). The optical fibers of the IO beam combiner could be connected in the MIRCx injection modules (represented on top of figure \ref{figmircx}) instead of the optical fibers of the standard MIRCx beam combiner. In the future, we plan to develop a new injection module which will allow to keep the optical fibers of both combiners connected. This will reduce the risks when handling the fibers.

\section{Fringe Tracker Software}
\label{ftsoft}
\subsection{Software Architecture}
The SPICA-FT software follows the CHARA compliant architecture, based on client/server architecture and GNOME ToolKit (GTK) Graphical User Interfaces (GUIs). It runs on the same machine (Linux Xubuntu operating system) than the MIRCx control command Software. The {\it credone\_control\_server} (MIRCx software) reads frames from the camera and writes them to a shared memory. The {\it mircx\_server} reads the frames from this shared memory and performs the real-time image processing. It writes the photometry of each beam, the correlated flux of each baseline and the associated errors into another shared memory.

The SPICA-FT software is fed by this last shared memory. For each frame, the first process, called {\it phase\_sensor}, reads the shared memory and computes the Phase Delay (PD), the Group Delay (GD), the PD SNR and the closure phases. Then these computed quantities are sent to the second process, called {\it opd\_controller}, which handles the fringe tracker state machine and computes the commands to be sent to the CHARA delay lines. The {\it opd\_controller} follows the algorithm described in next subsection.

\subsection{Algorithm}
%
%
The algorithm of SPICA-FT is mainly an adaptation of GRAVITY's algorithm presented in Lacour et al 2019\cite{lacour2019}. It is fed with the 15 complex coherent flux $\Gamma_{ij}$ of baselines $ij$ and the 6 photometries $F_i$ of telescopes $i$ estimated by the fringe sensor. It comes either from the standard MIRC-X combiner, the pairwise IO combiner of SPICA-FT, or MYSTIC combiner. These coherent flux enable the computation of several observables necessary for the generation of the control commands $u_i$ sent to the delay lines: the 15 phase-delays $\mathbf{\Phi}=\left(\Phi_{ij}\right)$ and group-delays $\mathbf{\Psi}=\left(\Psi_{ij}\right)$, the 20 closure phases $\mathbf{\Theta}=\left(\Theta_{ijk}\right)$ and an estimation of the variance of the phase-delay estimator ($Var\left(\Phi_{ij}\right)$), derived from the $\mathrm{SNR}\left(\Gamma_{ij}\right)$, for $(i,j,k)\in[\![1,6]\!]^3$ and $i<j<k$.

\subsection{Computation of the observables} \label{sec:observablecomputation}

\begin{figure}
    \centering
    \includegraphics[width=\linewidth]{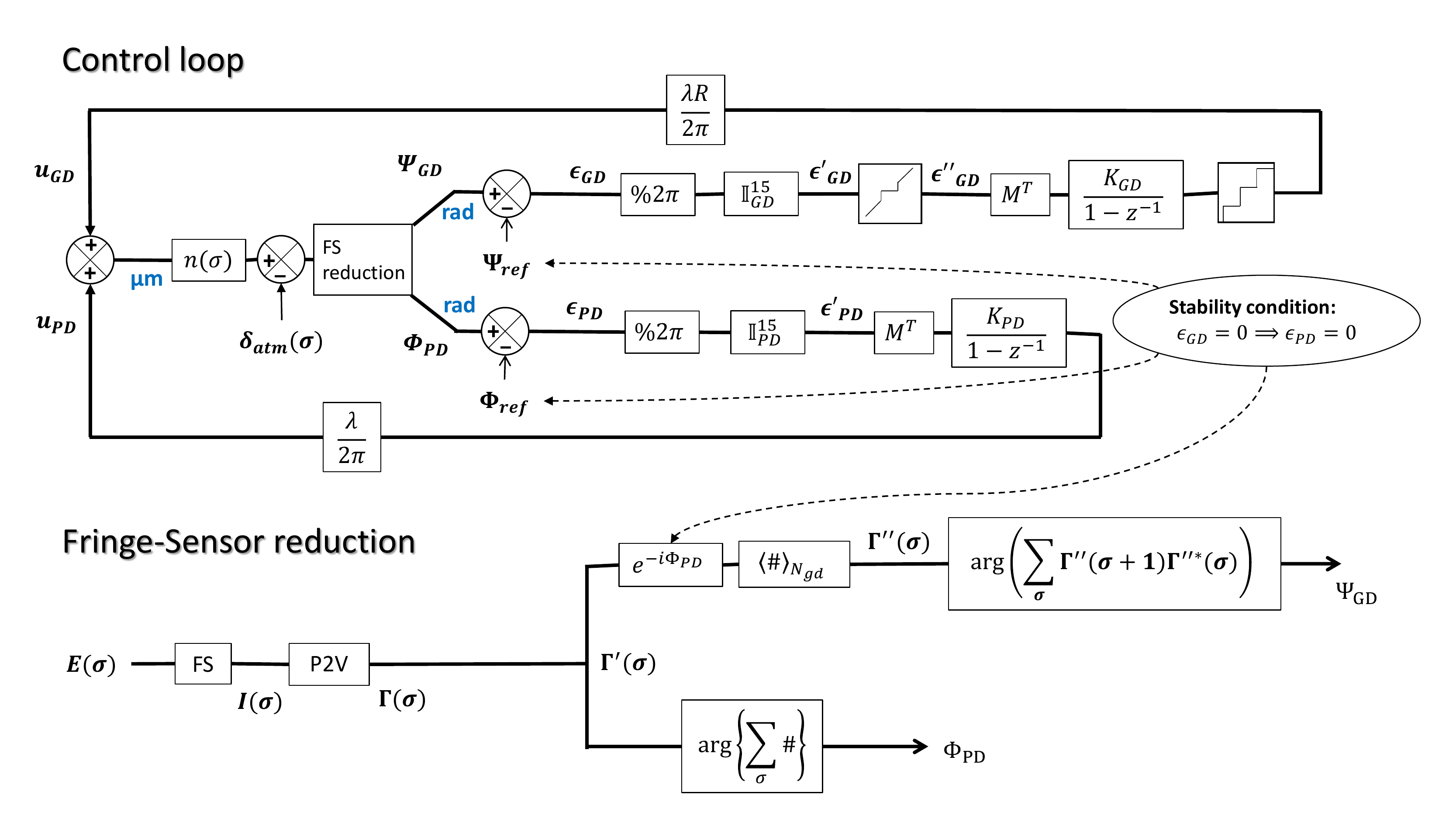}
    \caption{Block-diagram of the SPICA-FT TRACK state. For clarity, we don't show the closure phase computation that feed the reference vectors $\Phi_{ref}$ and $\Psi_{ref}$.}
    \label{fig:blocdiagram}
\end{figure}

As shown on the bottom part of the block-diagram in figure \ref{fig:blocdiagram}, PD and GD estimators are computed with different temporal frequencies. The phase-delays at time $n$ are computed using only the last coherent flux of all spectral channels,

\begin{equation}
    \Phi^n_{ij} = \arg{\left(\sum_\lambda^{N_\lambda}\Gamma'^n_{ij,\lambda}\right)},
\end{equation}
artificially reconstructing a white fringe from all spectral channels of the spectrograph. On the other hand, the group-delays are computed using the averaged coherent flux of each spectral channel on the $N_{gd}$ last frames corrected from the instantaneous phase-delays
\begin{equation}
\Gamma''^n_{ij,\lambda} = \left\langle\Gamma'_{ij,\lambda}\exp{\left(-i\Phi_{ij}\right)}\right\rangle_{N_{gd}}
\label{eq:cfgdlissé}
\end{equation}
and computing the phase difference of these averaged phasors between all neighbouring spectral channels:
\begin{equation}
\Psi^n_{ij} =\arg\left( \sum_{\lambda=1}^{N_\lambda-1} \Gamma''^n_{ij,\lambda+1} \Gamma''^{n*}_{ij,\lambda} \right).
\label{eq:gdlissé}
\end{equation}

$\mathbf{\Psi^n_{ij}}$ is not exactly the group-delays but only the phase differences proportional to the physical group-delays $\mathbf{GD}$ with the relation
\begin{equation}
    \mathbf{GD}=\dfrac{\bar{\lambda} R}{2\pi}\cdot\mathbf{\Psi},
\end{equation}
when expressed in microns, where $R$ and $\bar{\lambda}$ account for the spectral resolution and the mean wavelength of the instrument. From this, we see that the physical group-delays are defined modulo $\bar{\lambda} R$, i.e. a horizon $R$ times wider than the phase-delays which are defined, in microns, modulo $\bar{\lambda}$. The number of frames used for computing the group-delays must both increase enough the low SNR of the instantaneous group-delay estimator and remain shorter than the temporal scale of the drift of the fringe packet it compensates for. Acccording to Lawson et al 2000~\cite{lawson_least-squares_2000}, the GD estimator is expected to be 20 times noisier than PD estimator, suggesting an optimal averaging length $N_{gd}=20$ which corresponds to 80~ms at the fastest frame rate.

The closure phases of $\mathbf{\Phi}$ and $\mathbf{\Psi}$, expected to be independent from atmosphere, are computed on a longer temporal scale (typically $N_{cp}=300~frames$) to maximise their SNR and because they are expected to vary more slowly than $\mathbf{GD}$ and $\mathbf{PD}$ over time:
\begin{equation}
\Theta^{PD}_{ijk} =\arg\left( \left\langle\sum_{\lambda=1}^{N_\lambda} \Gamma'_{ij,\lambda} \sum_{\lambda=1}^{N_\lambda}\Gamma'_{jk,\lambda} \sum_{\lambda=1}^{N_\lambda}\Gamma'^*_{ik,\lambda} \right\rangle_{N_{cp}}\right)
\label{eq:cppd}
\end{equation}
and
\begin{equation}
\Theta^{GD}_{ijk} =\arg\left(\left\langle\Gamma^{(3)}_{ij} \Gamma^{(3)}_{jk} \Gamma^{(3)*}_{ik} \right\rangle_{N_{cp}}\right)
\label{eq:cpgd}
\end{equation}
where 
\begin{equation}
\Gamma^{(3)}_{ij} =\sum_{\lambda=1}^{N_\lambda-1} \Gamma''_{ij,\lambda+1} \Gamma''^{*}_{ij,\lambda}.
\label{eq:cfgd}
\end{equation}

An additional estimator at the heart of the SPICA-FT controller is the PD variance estimator. It is necessary for the controller in order to perform weighted least-square minimization of the PD and GD estimators and for its state-machine.
So far we tried two computations for the PD variance, the first one following the definition of Equation~14 of Lacour et al 2019\cite{lacour2019} and the second following the expression:

\begin{equation}
\sigma^2_\phi = \dfrac{\displaystyle\sum_\lambda \mathrm{Var}\left({\Re\left\lbrace\Gamma_\lambda\right\rbrace}\right)+\mathrm{Var}\left(\Im\left\lbrace\Gamma_\lambda\right\rbrace\right)}{2 \displaystyle\sum_\lambda \left\langle\left|\Gamma_{\lambda}\right|^2\right\rangle_{5DIT}}
\label{eq:varGD}
\end{equation}

which is more robust to PD jitters and thus seems better suited to fringe detection as we will see in Section.~\ref{results}.

\subsection{Computation of the commands}

\begin{figure}[h!]
    \centering
    \includegraphics[width=.75\linewidth]{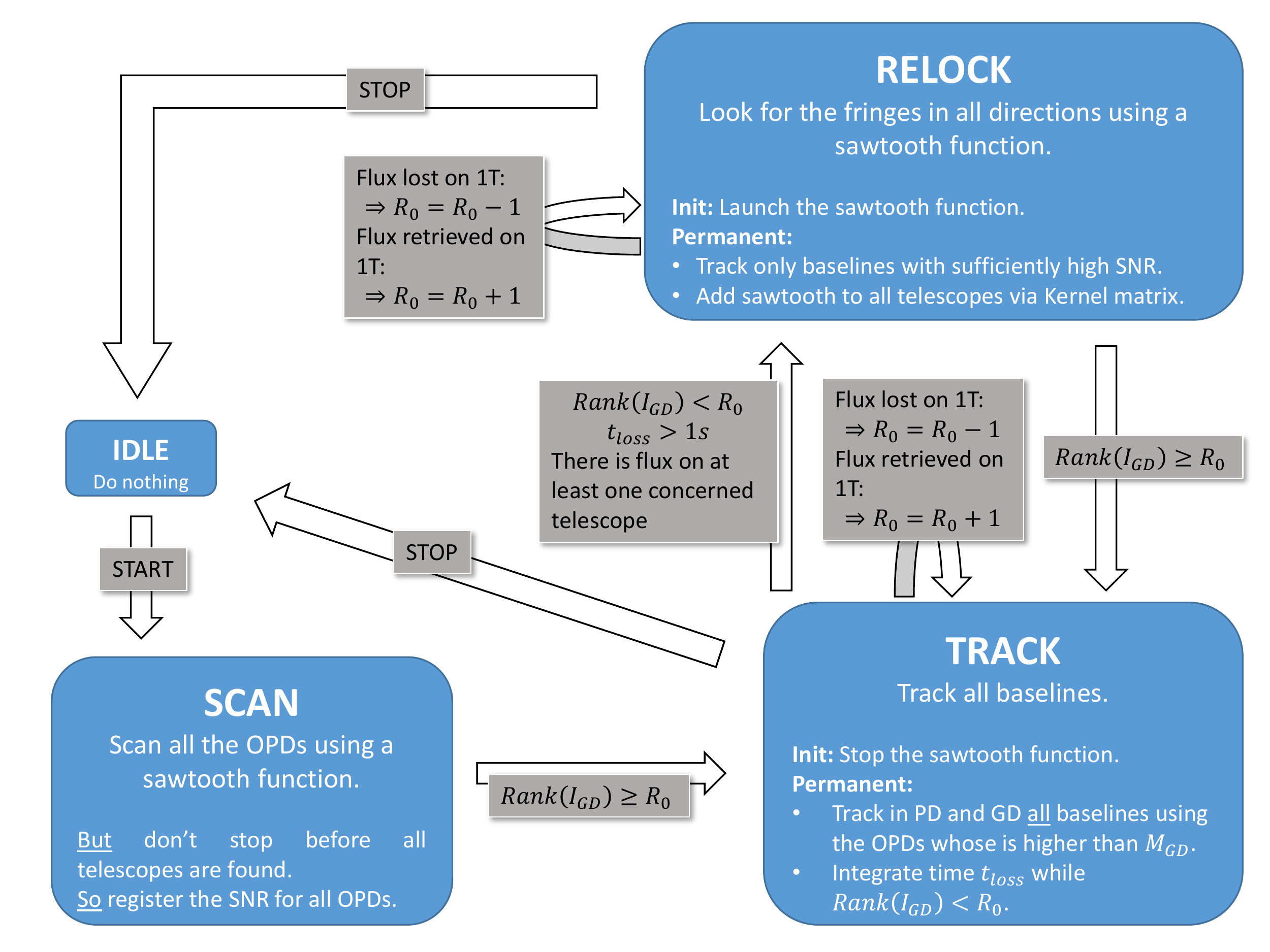}
    \caption{State-machine of SPICA-FT.}
    \label{fig:statemachine}
\end{figure}

The SPICA-FT fringe-tracking algorithm is based on the switch between two main states, TRACK and RELOCK. The latter is triggered each time flux is lost on one of the available telescopes. At the beginning of each observation, a specific third state, named SCAN, is triggered with the role of scanning the OPDs of all baselines to set the SNR thresholds parameters and find the fringes. This state-machine is schematized in figure \ref{fig:statemachine}.

During the permanent regime, TRACK and RELOCK controls are generated at each time $n$ such that the commands sent to the delay lines are
\begin{equation}
    \mathbf{u}^n = \mathbf{u}^n_{TRACK}+\mathbf{u}^n_{RELOCK}
\end{equation}
where both $\mathbf{u}^n_{TRACK}$ and $\mathbf{u}^n_{RELOCK}$ are null when respectively no telescopes and all telescopes are cophased and both are non-null when only a subset of telescopes are cophased.

\subsubsection{The TRACK state}

The control logic of the TRACK state is drawn as the upper block-diagram of figure \ref{fig:blocdiagram}. Since it is an adaptation to six telescopes of the GRAVITY fringe-tracking algorithm already explained in details in Lacour et al 2019\cite{lacour2019}, we focus here only on the implementations that differ from the article or that are of particular interest.

The definition of the reference vectors $\Phi_{ref}$ and $\Psi_{ref}$ is not straightforward. The role of these vectors is to handle the closure phases due to the instrument or to the astrophysical target. The closure phases make the measured OPDs inconsistent with pure atmospheric pistons. If we don't subtract it \textit{a priori}, they will thus spread over all OPDs when processing the least-square minimisation at the next step and the final projection into piston-space will be incorrect. For six telescopes, there are 20 closure phases of which 10 are independent. The correction is done by choosing 10 independent closure phases $\Theta_{ijk}$ and, for each one, subtracting it to the PD and GD estimators of one of its baselines. A simple way to choose 10 independent closure phases is to take all those which involve the same telescope. The reference vectors play thus this role of subtraction on the suitable baselines, using the slowly evolving estimated closure phases. To be robust to cases where some telescopes are not available, making impossible the measurement of the closure phases it belongs to, we defined six reference vectors

\begin{align}
	 \mathbf{\Phi_{ref}} = \begin{pmatrix}
           0 \\
           0 \\
           0 \\
           0 \\
           0 \\
           \Theta_{123} \\
           \Theta_{124} \\
           \Theta_{125} \\
           \Theta_{126} \\
           \Theta_{134} \\
           \Theta_{135} \\
           \Theta_{136} \\
           \Theta_{145} \\
           \Theta_{146} \\
           \Theta_{156} \\
         \end{pmatrix} & ,
         \begin{pmatrix}
           0 \\
           -\Theta_{123} \\
           -\Theta_{124} \\
           -\Theta_{125} \\
           -\Theta_{126} \\
           0 \\
           0 \\
           0 \\
           0 \\
           \Theta_{234} \\
           \Theta_{235} \\
           \Theta_{236} \\
           \Theta_{245} \\
           \Theta_{246} \\
           \Theta_{256} \\
         \end{pmatrix},
         \begin{pmatrix}
           \Theta_{123} \\
           0 \\
           -\Theta_{134} \\
           -\Theta_{135} \\
           -\Theta_{136} \\
           0 \\
           -\Theta_{234} \\
           -\Theta_{235} \\
           -\Theta_{236} \\
           0 \\
           0 \\
           0 \\
           \Theta_{345} \\
           \Theta_{346} \\
           \Theta_{356} \\
         \end{pmatrix},
         \begin{pmatrix}
           \Theta_{124} \\
           \Theta_{134} \\
           0 \\
           -\Theta_{145} \\
           -\Theta_{146} \\
           \Theta_{234} \\
           0 \\
           -\Theta_{245} \\
           -\Theta_{246}  \\
           0 \\
           -\Theta_{345} \\
           -\Theta_{346} \\
           0 \\
           0 \\
           \Theta_{456} \\
         \end{pmatrix},
         \begin{pmatrix}
           \Theta_{125} \\
           \Theta_{135} \\
           \Theta_{145} \\
           0 \\
           -\Theta_{156} \\
           \Theta_{235} \\
           \Theta_{245} \\
           0 \\
           -\Theta_{256} \\
           \Theta_{345} \\
           0 \\
           -\Theta_{356} \\
           0 \\
           -\Theta_{456} \\
           0 \\
         \end{pmatrix},
         \begin{pmatrix}
           \Theta_{126} \\
           \Theta_{135} \\
           \Theta_{146} \\
           \Theta_{156} \\
           0 \\
           \Theta_{236} \\
           \Theta_{245} \\
           \Theta_{256} \\
           0 \\
           \Theta_{346} \\
           \Theta_{356} \\
           0 \\
           \Theta_{456} \\
           0 \\
           0 \\
         \end{pmatrix}
         \label{eq:refvectors}
\end{align}

suitable to the situations where respectively each of the telescope 1, 2, 3, 4, 5 and 6 is cophased with the array. The GD reference vector $\mathbf{\Psi_{ref}}$ follows the same definition.

The PD and GD errors $\epsilon_{PD}$ and $\epsilon_{GD}$, compatible with piston-space projection, are finally obtained from the difference between the $\mathbf{\Phi}$ and $\mathbf{\Psi}$ estimators and their respective reference vectors $\mathbf{\Phi_{ref}}$ and $\mathbf{\Psi_{ref}}$. 

From the values of PD variances of all baselines provided by the first stage of the fringe-tracker, we compute the matrices $\mathbb{I}^{15}_{PD}$ and $\mathbb{I}^{15}_{GD}$ which play the roles of:
\begin{itemize}
    \item estimating, by a weighted least-square minimization, 15 new $\epsilon'_{PD}$ and $\epsilon'_{GD}$ from the 15 PD and GD errors computed by the first stage of the fringe-tracker: the baselines whose SNR is lower than $\mathrm{SNR}_{GD}$ are not taken into account whereas the others are basically weighted by their SNR. $SNR_{GD}$ is set by the operator or determined by the state SCAN at the beginning of the observation.
    \item filtrating the PD commands on pistons for which the variance is estimated lower than a given threshold, meaning this command would have no physical origin.
\end{itemize}

Then the GD errors go through an extra step that nulls the errors whose absolute value is lower than a given threshold, currently fixed to $\lambda$, to prevent the two controls PD (that, because of its wrapping over $2\pi$, corrects only absolute errors lower than $\lambda/2$) and GD to send twice the same command to the delay lines.
These errors are then projected into piston-space thanks to the pseudo-inverse of the matrix of projection from piston to OPDs:
\setcounter{MaxMatrixCols}{20}
\begin{equation}
    \mathbf{M} = 
    \begin{pmatrix}
    1&1&1&1&1&0&0&0&0&0&0&0&0&0&0 \\
    -1&0&0&0&0&1&1&1&1&0&0&0&0&0&0\\
    0&-1&0&0&0&-1&0&0&0&1&1&1&0&0&0\\
    0&0&-1&0&0&0&-1&0&0&-1&0&0&1&1&0\\
    0&0&0&-1&0&0&0&-1&0&0&-1&0&-1&0&1\\
    0&0&0&0&-1&0&0&0&-1&0&0&-1&0&-1&-1
    \end{pmatrix}^T
\end{equation}
to get the pistons errors

\begin{equation}
    \mathbf{\epsilon}_{p,PD} = \mathbf{M}^\dag \cdot \mathbf{\epsilon}'_{PD}
    \label{eq:pistonsPD}
\end{equation}
and
\begin{equation}
    \mathbf{\epsilon}_{p,GD} = \mathbf{M}^\dag \cdot \mathbf{\epsilon}''_{GD}
    \label{eq:pistonsGD}
\end{equation}

Finally, at each time $n$, the controllers PD and GD are integrated following the classical PI controller equations
\begin{equation}
    \mathbf{u}_{PD}^n = \mathbf{u}_{PD}^{n-1} + K_{PD}\cdot \mathbf{\epsilon}_{p,PD}^n
    \label{eq:PDcommand}
\end{equation}
and
\begin{equation}
    \mathbf{u}_{GD}^n = \mathbf{u}_{GD}^{n-1} + K_{GD}\cdot\mathbf{\epsilon}_{p,GD}^n
    \label{eq:GDcommand}
\end{equation}
before the GD control go again through an extra step that converts the piston commands into an integer number of the mean wavelength $\bar{\lambda}$ of the fringe-sensor to prevent from blurring the fringes.

Thus, the total command from the TRACK state is:
\begin{equation}
    \mathbf{u}^n_{TRACK} = \mathbf{u}^n_{GD} + \mathbf{u}^n_{PD}
    \label{eq:trackcommand}
\end{equation}

\subsubsection{The RELOCK state}

The fringe-tracker enters the RELOCK state when the SNR on all baselines involving a given telescope are below the threshold $\mathrm{SNR}_{GD}$, leading to the reduction of the rank of the matrix $\mathbb{I}^{15}_{GD}$ by as much unities as lost telescopes, and remains below the threshold during at least one second. An additional condition for triggering the RELOCK state is that flux is present on the lost telescopes, to avoid looking for unavailable telescopes. This state is of a major importance on SPICA-FT which is a fiber-fed spectro-interferometer. Injection losses can occurs often, all the more so as there are six fibers to feed with six different adaptive optic systems with their own weaknesses. 

It is worth mentioning that during the RELOCK state, the GD and PD control loops keep generating $\mathbf{u}_{TRACK}$ commands for cophasing as much telescopes as possible. When it starts, the RELOCK state launches a \textit{sawtooth} function that will be used to generate additional commands on the telescopes. The RELOCK command sent to all telescopes is computed as follows:

\begin{equation}
    \mathbf{u}^n_{RELOCK} = \mathbb{K}^{6,n} \cdot \mathbf{v}^T \cdot u^n_{saw}
\end{equation}
where the multiplication of the vector velocities $\mathbf{v}^T$ by the matrix
\begin{equation}
    \mathbb{K}^{6,n}=\mathbf{I}^6-\mathbb{I}^{6,n}_{GD}
    \label{eq:}
\end{equation}
with $\mathbb{I}^{6,n}_{GD} = \mathbf{M}\cdot \mathbb{I}^{15,n}_{GD}\cdot \mathbf{M}^T$, works such that:
\begin{itemize}
    \item the components of the vector $\mathbf{v}$ of lost telescopes remains unchanged,
    \item the components of the vector $\mathbf{v}$ corresponding to telescopes belonging to a same cophased group are added up together.
\end{itemize} 
Doing so, the cophased telescopes are not disturbed by the RELOCK command while the lost telescopes are scanning. However, this implementation works only with specific velocity vectors $\mathbf{v}$ non redundant in a sense that for any subset of cophased telescopes, the addition of its components never equals the computed velocity of another subset (one or more) of the complementary telescopes. This is achieved with the vector
\begin{equation}
      \mathbf{v}=\begin{pmatrix}
-8.25& -7.25& -4.25& 1.75& 3.75& 8.75 
\end{pmatrix}^T
\label{eq:vvector}
\end{equation}
except for one very specific situation: when at the same time the telescopes 1, 3 and 4 are cophased together while the telescopes 2, 5 and 6 are also cophased together. If this occurs, we only need to change the sign of the velocity of one of the two groups to keep searching.

The \textit{sawtooth} function triggered at the beginning of RELOCK state is designed to scan around zero, alternating in both directions with a little overlap at each change, while gradually moving away faster and faster. Its behaviour is visible on figure 5 of Lacour et al 2019\cite{lacour2019}.

\subsubsection{The SCAN state}

The SCAN state is specific to the beginning of an observation. After the delay lines are set up to the positions where the coherence of the telescopes are expected according to the coordinates of the star and of the stellar interferometer, they receive new commands

\begin{equation}
    \mathbf{u}_{SCAN} = \mathbf{v_0}\cdot u_{saw},
\end{equation}
where $u_{saw}$ is the same \textit{sawtooth} function than for RELOCK state and
\begin{equation}
    \mathbf{v_0} = \begin{pmatrix} 0&1&2&3&4&5\end{pmatrix}^T
\end{equation} is a velocity vector chosen to scan all telescopes except the reference one (here the first one) around the start positions while recording the minimal SNR, corresponding to the absence of fringes, and maximal SNR, corresponding to the center of the fringe packet. The positions where the null GD is found between the reference telescope and the five others is also recorded in order to send the telescopes to these positions at the end of the SCAN state.

For each baseline, the records of the minimal and maximal SNRs enable to set the PD and GD thresholds $\mathrm{SNR}_{PD}$ and $\mathrm{SNR}_{GD}$, two parameters of high importance for the RELOCK and TRACK states. These thresholds can differ from one baseline to another since the visibilities depend on the baselength and the injection conditions can be very different depending on the telescope.

\section{Results on Sky} \label{results}
\subsection{Phase-tracking performance}

\begin{figure}
    \centering
    \includegraphics[width=\linewidth]{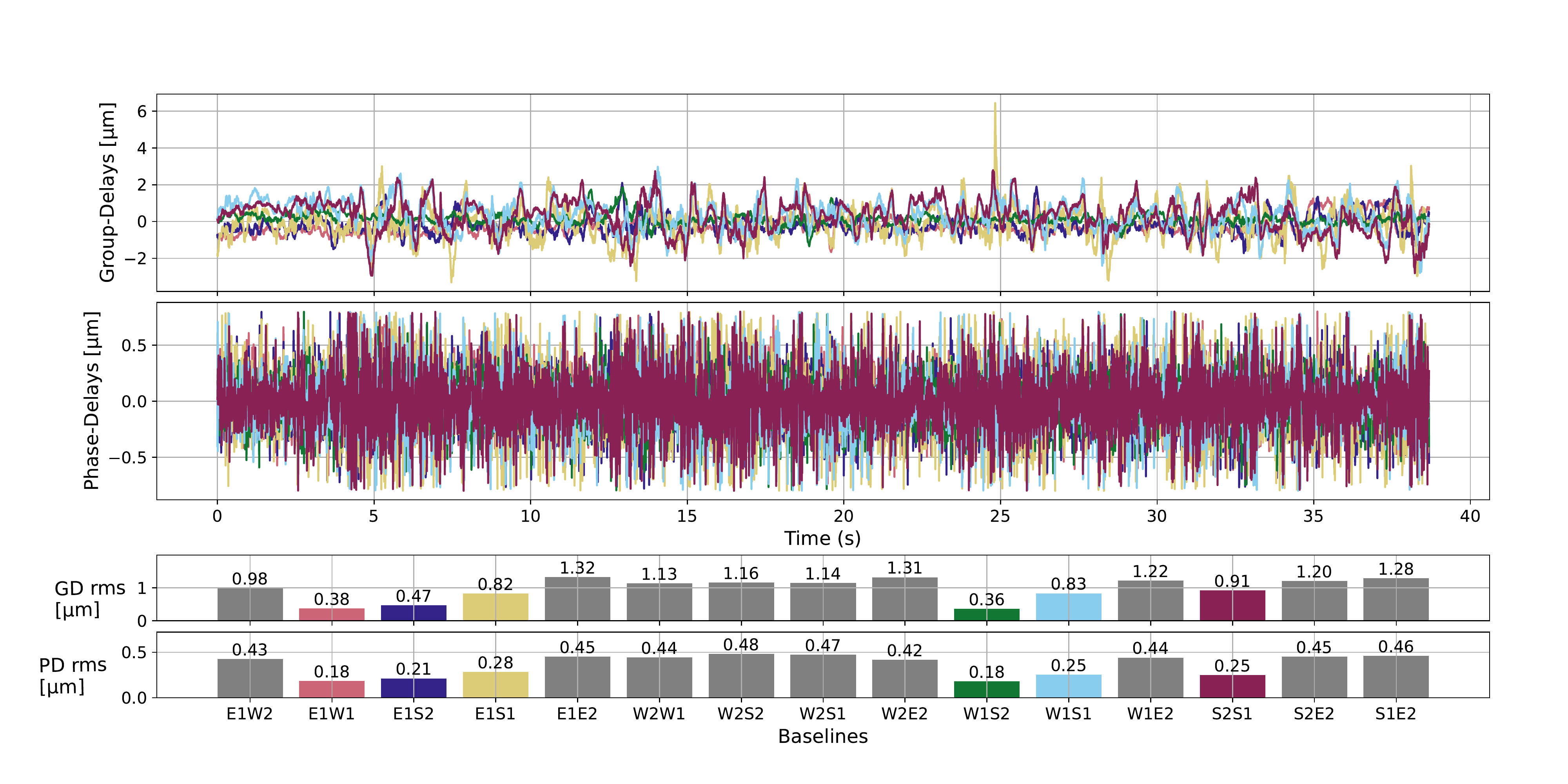}
    \caption{Phase-tracking with E1S1S2W1 on the star HD186155, magnitude 4 and diameter $\o_H\simeq0.58$~mas leading to $\left|V_H\right|\simeq0.4$ on the longest baseline E1S1.}
    \label{fig:phasetracking4T}
\end{figure}

On figure \ref{fig:phasetracking4T}, we show a $\sim38.5$~seconds sequence of phase-tracking with four telescopes (E1, S1, S2, W1) that occurred during our last commissioning run in May 2022. The tracked star is HD186155, a star of magnitude $m_H\simeq4.2$ and diameter $\o_H\simeq0.58$~mas. The visibility of the star is between 0.4 and 0.6 at 1.6~\micron for all involved baselines but the shortest one S1S2 which barely resolves it. During this sequence, SPICA-FT reached PD residuals below 200~nm rms on the baselines of the triangle E1S2W1 and 250~nm rms on the other baselines. The GD estimator remains around 0 with standard-deviation of about 400~nm for the triangle E1S2W1 and 800~nm on the three other baselines.
The PD estimator of non-tracked baselines are all showing PD estimator rms between 420 and 460~nm.

\subsection{Limitations and expected improvements}

\subsubsection{Fringe jumps}

\begin{figure}
    \centering
    \includegraphics[width=\linewidth]{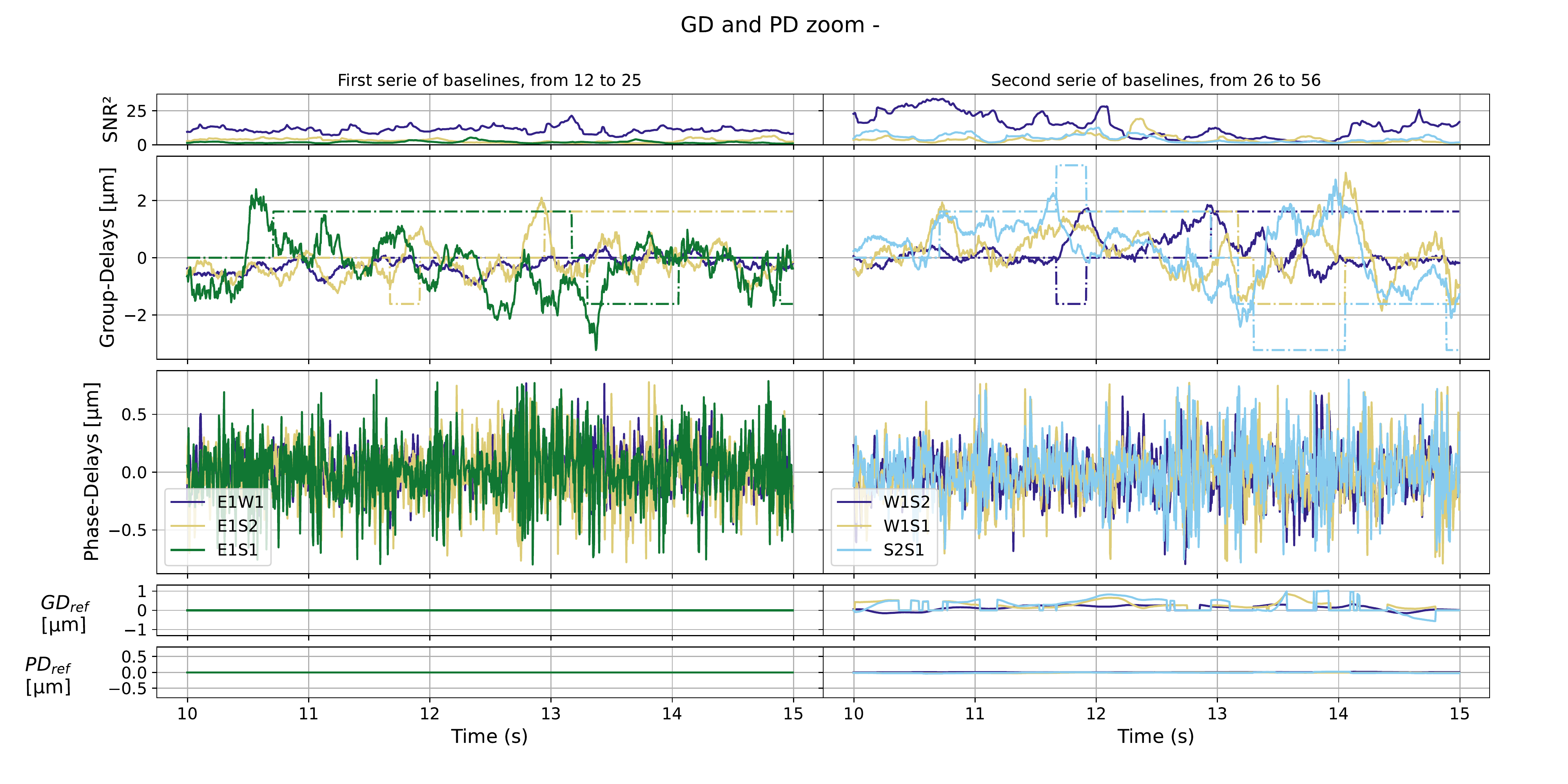}
    \caption{Zoom on the phase-tracking sequence with E1S1S2W1.}
    \label{fig:fringejumps}
\end{figure}

Although the phase-delay residuals were low, we experienced during the fringe-tracking more fringe jumps that we expect for. This is illustrated on figure \ref{fig:fringejumps} with a zoom on the phase-tracking sequence already shown in figure \ref{fig:phasetracking4T}. Their frequency compared to the dispersion evolution time scale is too high to be explained by the normal functioning of the GD loop. Fringe jumps can occur when the OPD moves of more than a wavelength, either very quickly compared to the PD loop latency or during a loss of flux that prevents the PD loop to see the shift. But the observing conditions were sufficiently good to put aside this explanation, at least considering such a high frequency of fringe jump events.

We thus looked more precisely at the behaviour of the GD reference vector, appearing at the bottom of figure \ref{fig:fringejumps}, which give the targets on all baselines for the GD control loop and we saw that it is far noisier than the PD reference vector. As a consequence, the implementation that we chose, putting the component of the reference vector to zero as soon as its SNR is too low, implies a lot of resets to zero which could explain such an instability for the GD control loop. The plan now is to test a phase-tracking either without GD reference vector or with a different implementation of it.

We believe though that a smarter implementation of the GD reference vector will lead to a substantial reduction of the fringe jumps frequency.

\subsubsection{SNR estimation and fringe detection}

\begin{figure}[ht!] \centering
\begin{subfigure}[b]{\linewidth}\centering
\includegraphics[width=0.6\textwidth]{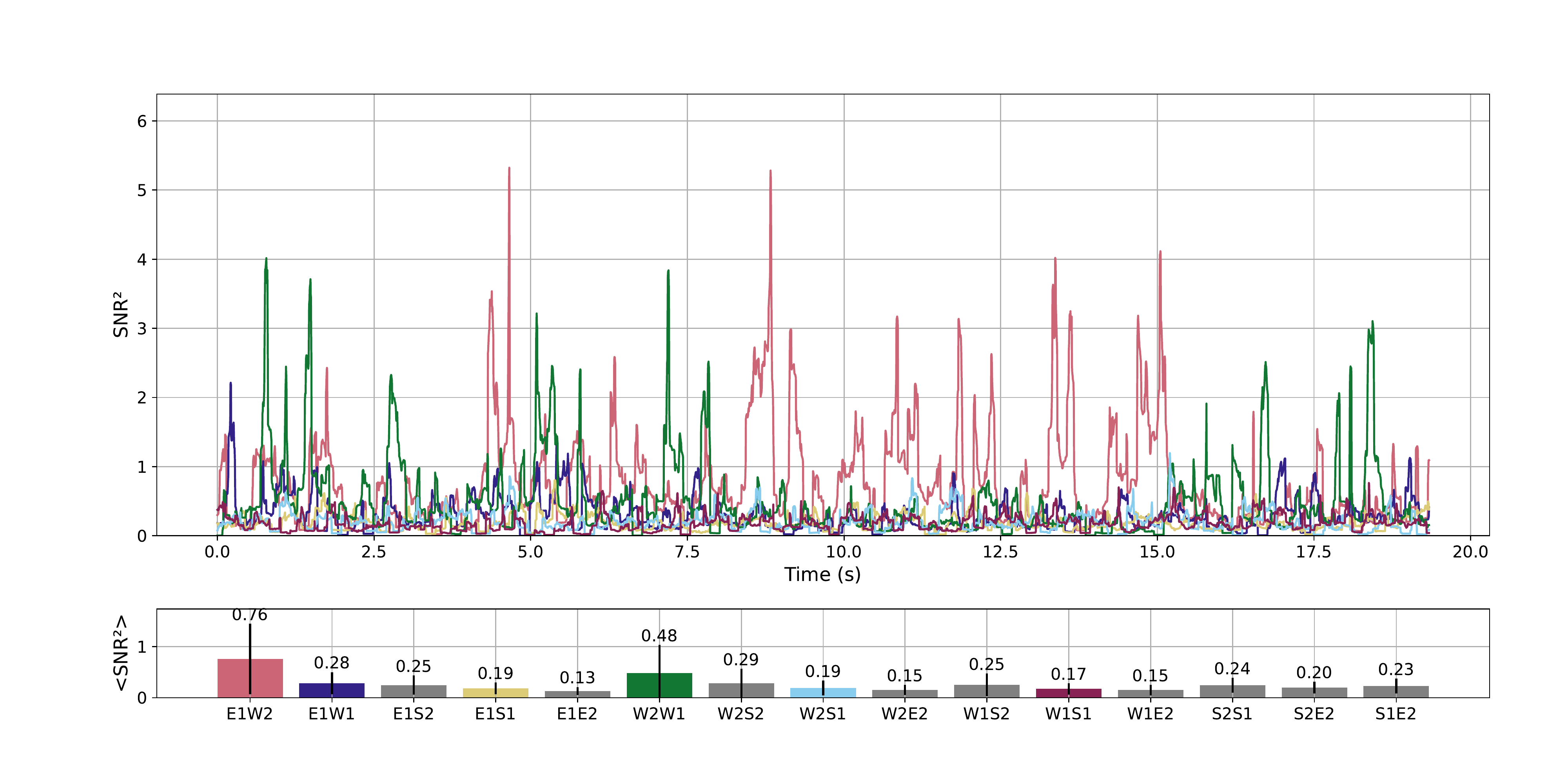}
\caption{Squared SNR of phase-delay estimator averaged over 40 frames, using Equation~14 of Lacour et al 2019\cite{lacour2019}.}
\label{subfig:spicaft_pdvariance}
\end{subfigure}
\begin{subfigure}[b]{\linewidth}\centering
\includegraphics[width=0.6\textwidth]{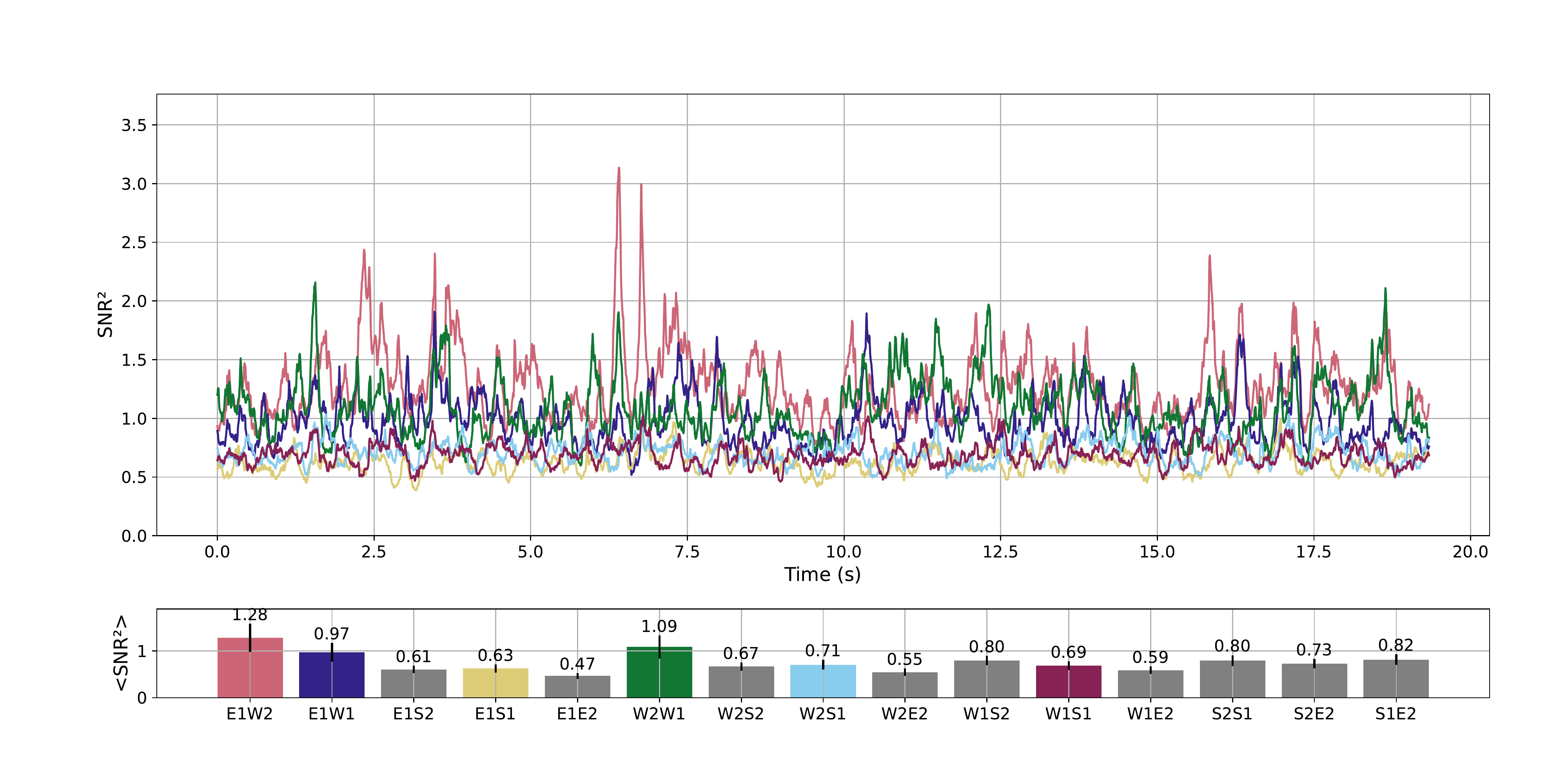}
\caption{Squared SNR of phase-delay estimator averaged over 40 frames, using Equation~\ref{eq:varGD}.}
\label{subfig:spicaft_gdvariance}
\end{subfigure}
\caption{Comparison between the estimation of squared SNR relying on the integration of coherent fluxes over 5~DIT and the one relying on the incoherent addition of coherent flux modules tried during the commissioning of May 2022. During the computation, MIRC-X was group-tracking the telescopes E1, W1 and W2.}
\label{fig:spicaft_snrestimation}
\end{figure}

Another source of improvement of SPICA-FT will be on the estimation and usage of the SNR of the measurements. We already defined in section~\ref{sec:observablecomputation} the two ways of computing the SNR of the PD estimator. figure \ref{fig:spicaft_snrestimation} shows the difference of behaviour of these two estimators during a sequence when MIRC-X is group-tracking only on the telescopes E1, W1 and W2. We plot on these figures the curves concerning also the telescope S1 for comparison with the tracked baselines. The fact that MIRC-X is only group-tracking implies that the fringes are still moving a lot. But we were monitoring during the sequence recording that the fringes were always present. The integration time was 20~ms, longer than the atmospheric coherence time at this moment ($\sim10$~ms). The estimator relying on the coherent integration often reaches very high values, which is a signal for the presence of fringes. However, it also falls down very often to values corresponding to the absence of fringes. The error bars in the bar diagram show that the rms value of this estimator is very high, making it poorly reliable to detect jittering fringes. Shortening the integration time to 8~ms didn't show improvement in this regard. On the other side, the estimator summing up the modules of coherent flux shows lower values in presence of fringes but also a lower standard deviation, which as a whole seem to make it more suitable for detecting jittering fringes. For this reason, we will keep this estimator for the moment.

On figure \ref{subfig:spicaft_gdvariance}, we also see that the estimation of the SNR on the non-tracked baselines, where there is no fringe, can be rather different from a baseline to another (0.47 on E1E2 and 0.82 on S1E2). This suggests that defining one threshold per baseline will improve the state-machine of SPICA-FT.

\section{Conclusion}
We report on the first on-sky tests of the SPICA-FT fringe tracker installed at the CHARA array. During an observing run in May 2022, we stabilized the fringes of 4 telescopes with an OPD residuals around 200nm during several minutes on several stars with magnitude between 4 and 4.5 in H band.

We identify several limitations in the fringe tracker software that could be easily overcome in the near future. We would like to reduce substantially the fringe jumps and reach OPD residuals of 100~nm with 6 telescopes. In addition to the software improvements, we plan to install our IO beam combiner at the entrance of the MIRCx spectrograph. Before this installation, we will develop a new fiber injection module which will allow to switch easily between the MIRCx and SPICA-FT fibers. 

We think SPICA-FT (IO beam combiner + Software) will be ready by the end of 2022. In the next 6 months, we will derive its full performance in terms of sensitivity and OPD residuals.

\bibliography{spicaft} 
\bibliographystyle{spiebib} 

\end{document}